%% file: main.tex
\documentclass[journal]{IEEEtran}

\usepackage[T1]{fontenc}%

\usepackage{cite}

\ifCLASSINFOpdf
 \usepackage[pdftex]{graphicx}
\else
\fi

\usepackage{pgfplots}
\pgfplotsset{compat=newest}

\usepackage{amsmath}
\usepackage{amsthm}
\usepackage{amssymb}
\usepackage{mathrsfs}
\usepackage{balance}

\usepackage{bm}

\usepackage[capitalize]{cleveref}
\crefname{equation}{\unskip}{\unskip}

\newcommand{\CSSS}{\mathscr{C}_{\mathsf{SSS}}}

\newcommand{\by}{\boldsymbol{y}}
\newcommand{\emax}{e_\mathsf{max}}
\newcommand{\w}{\boldsymbol{W}}
\newcommand{\bC}{\boldsymbol{C}}

\newcommand{\x}{\boldsymbol{x}}
\newcommand{\s}{\boldsymbol{s}}
\newcommand{\set}[1]{\mathcal{#1}}
\newcommand{\design}[1]{\mathfrak{#1}}
\newcommand*{\Scale}[2][4]{\scalebox{#1}{\ensuremath{#2}}}%

\newcommand{\code}{\mathscr{C}}

\newtheorem{theorem}{Theorem}

\newtheorem{corollary}{Corollary}
\newtheorem{example}{Example}
\newtheorem{remark}{Remark}

\definecolor{mycolor2}{RGB}{170,15,69}
\definecolor{mycolor3}{RGB}{229,121,80}
\definecolor{mycolor1}{RGB}{35,35,35}
\definecolor{mycolor4}{RGB}{251,165,93}
\definecolor{mycolor5}{RGB}{100,192,165}
\definecolor{mycolor6}{RGB}{56,128,164}
\definecolor{mycolor7}{RGB}{94,79,162}

\definecolor{TUMGreen}{rgb}{0.57,0.67,0.42}

\begin{document}

\title{Privacy-Preserving Coded Mobile Edge Computing for Low-Latency Distributed Inference}

\author{Reent~Schlegel,~\IEEEmembership{Student Member,~IEEE},
        Siddhartha~Kumar,
        Eirik~Rosnes,~\IEEEmembership{Senior Member,~IEEE},
        and~Alexandre~Graell~i~Amat,~\IEEEmembership{Senior Member,~IEEE}%
        \thanks{This work was financially supported by the Swedish Research Council under grant 2020-03687. This paper was presented in part at the IEEE Global Communications Conference (GLOBECOM), Taipei, Taiwan, December 2020 \cite{Schlegel}.}
\thanks{R. Schlegel, S. Kumar, and E. Rosnes are with Simula UiB, Bergen, Norway, e-mail: \{reent,~kumarsi,~eirikrosnes\}@simula.no.}%
\thanks{A. Graell i Amat is with the Department of Electrical Engineering, Chalmers University of Technology, Gothenburg, Sweden, e-mail: alexandre.graell@chalmers.se, and with Simula UiB, Bergen, Norway.}}%

\maketitle

\bstctlcite{IEEEexample:BSTcontrol} %
\begin{abstract}
We consider a mobile edge computing scenario where a number of devices want to perform a linear inference $\w\x$ on some local data $\x$ given a network-side matrix $\w$. The computation is performed at the network edge over  a number of edge servers. We propose a coding scheme that provides information-theoretic privacy against $z$ colluding (honest-but-curious) edge servers, while minimizing the overall latency\textemdash comprising upload, computation, download, and decoding latency\textemdash in the presence of straggling servers. The proposed scheme exploits Shamir's secret sharing to yield  data privacy and straggler mitigation, combined with replication to provide spatial diversity for the download. We also propose two variants of the scheme that further reduce latency. %
For a considered scenario with $9$ edge servers, the proposed scheme reduces the latency by $8\%$ compared to  the nonprivate scheme recently introduced by  Zhang and Simeone, while providing privacy against an honest-but-curious edge server. %

\end{abstract}


\IEEEpeerreviewmaketitle

\input{Introduction}
\input{Systemmodel}

\input{PrivateDistributedLinearInference}
\input{CommunicationAndComputationScheduling}

\input{Variants}	
\input{OptimizationAndNumericalResults}
\input{Conclusion}

\input{Appendix}

\ifCLASSOPTIONcaptionsoff
  \newpage
\fi

\balance 
\bibliographystyle{IEEEtran}
\bibliography{bibliography.bib}

\end{document}

%% file: Introduction.tex
\section{Introduction}

Mobile edge computing is a key enabler of delay-critical internet-of-things applications that rely on large data computing services \cite{Mach}, and has become a pillar of the 5G mobile network \cite{ETSI}. Offloading computations to far-away cloud services can be infeasible due to bandwidth constraints on the backhaul network and possibly large communication latency \cite{Mach}. To circumvent these shortcomings, the edge computing paradigm moves the computation power closer to the devices generating the data.%

Distributing computations over a number of servers at the edge of the wireless network  leads to major challenges, among them the presence of \emph{straggling} servers\textemdash the computation latency is dominated by the slowest server. The straggler problem has been addressed in the neighboring field of distributed computing in data centers by means of coding  \cite{Li,Lee,Albin1,Albin2,Yu,Reisizadeh,Dutta,Dutta2,Tandon,Karakus,Mallick}. The key idea in distributed computing is to introduce redundant computations across servers via an erasure correcting code such that the results from a subset of the servers is sufficient to recover (decode) the desired computation. Hence, the latency is no longer dominated by the slowest servers. Maximum distance separable (MDS) codes have been shown to provide excellent straggler resiliency \cite{Li,Lee}. Most works on coded computing neglect the impact of the decoding complexity on the latency.
An exception is \cite{Albin1, Albin2}, where it
was shown that the decoding latency may severely impact
the overall latency. Long MDS codes, in particular, entail a high decoding complexity, which may impair the overall latency.

In edge computing, besides the straggler problem, the incurred latency of uploading and downloading data through the wireless links is a genuine problem. To reduce the communication latency, in \cite{Tao,TaoStudy} subtasks are replicated across edge servers to introduce spatial diversity such that the edge servers can utilize zero-forcing precoding and serve multiple users simultaneously.  More recently, the authors in \cite{Osvaldo,Kuikui,LiTao} proposed to combine subtask replication for spatial diversity with an MDS code for straggler mitigation, borrowing the coding ideas from distributed computing. These works, however, neglect the latency entailed by the decoding operation. In \cite{Fri21}, a scheme combining rateless codes with irregular repetition was proposed, yielding significantly lower   latency (comprising the decoding latency)   than  the scheme in \cite{Osvaldo,Kuikui,LiTao}.

Performing computations over  
possibly untrustworthy edge servers raises also  privacy concerns.  The problem of user data privacy in the context of distributed computing in data centers in the presence of
stragglers has been addressed in, e.g., \cite{SalimStair,SalimRateless,YuLi,Yang}. The underlying  idea in these works is to utilize some form of secret sharing, i.e., encode the confidential user data together with random data such that small subsets of servers do not gain information about the confidential data. %

In this paper, %
we consider a similar scenario to the one in \cite{Osvaldo,Fri21} where multiple users wish to perform a linear inference $\w\x$ on some local data $\x$ given a network-side  matrix $\w$. Such operations arise in, e.g., recommender systems based on collaborative filtering, %
like  a shopping center application providing product recommendations and corresponding price offers\cite{Felferning}. Each customer has its  preferences, which are encoded by an attribute vector $\bm x$. Based on a  customer's preferences, the application  recommends products by mapping from a customer's preference (vector $\bm x$) to the likelihood that he/she would enjoy a given product via the system matrix $\bm W$.
For this scenario, we present a coding scheme that guarantees information-theoretic user data privacy against $z$  compromised (honest-but-curious) edge servers that collaborate to infer users' data,  while minimizing the incurred overall latency\textemdash comprising  upload, computation, download, and decoding latency.  %
The proposed scheme is based on Shamir's secret sharing (SSS) scheme \cite{Shamir} to achieve  data privacy as well as straggler mitigation\textemdash thereby reducing computation latency\textemdash combined with replication of  subtasks across multiple edge servers to allow for spatial diversity and joint beamforming (by means of zero-forcing precoding) in the download to reduce communication latency. A key feature of the proposed scheme is that, unlike the existing (nonprivate) schemes for straggler mitigation in edge computing \cite{Osvaldo,Kuikui,LiTao}, redundancy is introduced on the users' data\textemdash which enables privacy\textemdash instead of on the network-side matrix $\w$.

We also introduce two variants of  the scheme that further reduce latency. First, we note that the download phase can be performed simultaneously to the computation phase once the upload phase is completed, which reduces the overall latency especially when the communication cost is relatively high compared to the computation cost. To exploit this, we introduce a priority queue to determine the order  in which partial results should be downloaded from the edge servers. Second, we introduce an additional level of coding on $\w$. We show that the combination of the SSS code and the code on $\w$ results in a product code over intermediate results. %
Decoding can then be performed iteratively, iterating between the row and column component decoders. 

The proposed scheme entails an inherent tradeoff between computation latency due to straggling servers, communication latency, and user data privacy.
Interestingly, for a considered scenario with $9$ edge servers, the proposed scheme reduces the latency by  $8\%$ compared to the nonprivate scheme in \cite{Osvaldo}, while providing privacy against a single edge server. This somewhat surprising result is explained by the   high  decoding  complexity of the  scheme  in  \cite{Osvaldo} due to the use of a long MDS code (on $\w$), while the proposed schemes rely on short codes over both users' data and $\w$. Higher privacy levels can be achieved 
at the expense of higher latency. Furthermore, the additional coding on $\w$ significantly reduces the variance of the latency, which, for a scenario where the linear inference needs to be performed within a deadline, increases the probability of meeting the deadline.

\textit{Notation:} Vectors and matrices are written in lowercase and uppercase bold letters, respectively, e.g.,  $\boldsymbol{a}$ and $\boldsymbol{A}$, and all vectors are represented as column vectors.
The transpose of vectors and matrices is denoted by $(\cdot)^\top$. $\text{GF}(q)$ denotes the finite field of order $q$ and $\mathbb{N}$ denotes the positive integers. We use the notation $[a]$ to represent  the set of integers $\{1,2,\ldots,a\}$. Furthermore, $\left\lceil a/b\right\rceil$ is the smallest integer larger than or equal to $a/b$ and $\left\lfloor a/b\right\rfloor$ is the largest integer smaller than or equal to $a/b$. We represent permutations in cycle notation, e.g., the permutation $\pi = (1\;3\;2\;4)$ maps $1\mapsto3$, $3\mapsto2$, $2\mapsto4$, and $4\mapsto1$. In addition, $\pi(i)$ is the image of $i$ under $\pi$, e.g., $\pi(1) = 3$. Applying $\pi$ recursively $i$ times is denoted by $\pi^i$, e.g., $\pi^2(3) = 4$, and $\pi^0$ is an identity, e.g., $\pi^0(2) = 2$. The expected value of a random variable $X$ is denoted by $\mathbb{E}[X]$.

%% file: Systemmodel.tex
\section{System Model}\label{Sec: SystemModel}

\begin{figure}
		\centering
		\includegraphics[keepaspectratio, width = 1\columnwidth]{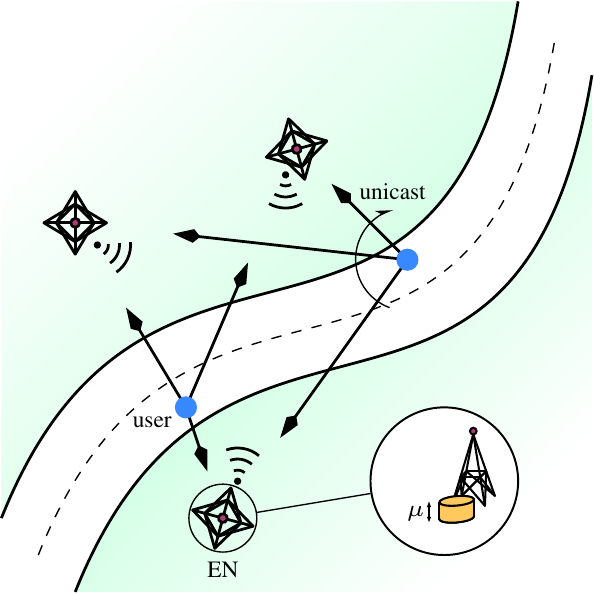}
		\caption{A mobile edge computing network with two users and three ENs.\label{fig:system_model}}
	\end{figure}

We consider a scenario with   $u$  single-antenna users, $\mathsf{u}_1, \ldots ,\mathsf{u}_{u}$, each wanting to compute the linear inference operation $\by_i=\w\x_i$ on its local and private data  $\x_i = (x_{i,1},\ldots,x_{i,r})^\top\in\mathrm{GF}{(q)}^r$ for some network-side public matrix $\w\in\textrm{GF}{(q)}^{m\times r}$. The operation is offloaded to the edge and is performed in a distributed fashion over a number of edge servers\textemdash hereafter referred to as edge nodes (ENs). We assume that there are $\emax$ ENs available at the network edge, and that the linear inference is performed over  $e\leq \emax$ ENs, where $e$ can be optimized. The $e$ ENs that perform the computation tasks are denoted $\mathsf{e}_1,\ldots,\mathsf{e}_{e}$. Each EN has a storage capacity $\mu$, $0<\mu \leq 1$, which is the fraction of $\w$ each EN can store, i.e., each EN can store up to $\mu mr$ elements from $\textrm{GF}{(q)}$. We assume that  $\w$ stays constant over a sufficient amount of time so that it can be stored on the ENs offline.  %
The system model is depicted in \cref{fig:system_model}.
    
	\subsection{Computation Runtime Model}
	\label{sec:RunTime}
	
	The ENs are in general multi-task nodes, may run several applications in parallel, and need to serve many users. As a result, they may straggle. We model this behavior with a random setup time $\lambda_j$ for each EN $\mathsf{e}_j$. The setup time is the time it takes an EN to start the computation after it received all the necessary data. Here, we assume the widely-adopted model in which the setup times are independent and identically distributed (i.i.d.) and modeled by an exponential distribution with parameter $\eta$, such that $\mathbb{E}[\lambda_j] = 1/\eta$  \cite{Osvaldo,Dean,Mallick}. Once set up, an EN needs $\tau$ time units to compute an inner product in $\text{GF}(q)^r$ for each of the users, i.e., it takes an EN $\tau$ time units to do $r$ multiplications and $r-1$ additions for all users. Consequently, to compute $d$ inner products for each user ($u d$ inner products in total), EN $\mathsf{e}_j$  incurs a latency of
		\begin{equation*}
			 \lambda_{j} + d\tau\,.
	\end{equation*}
	We define the \emph{normalized computation} latency of EN $\mathsf{e}_j$ (normalized by $\tau$) as
	\begin{equation*}
		{\mathsf{L}}^\mathsf{comp}_{j} =  \frac{\lambda_{j}}{\tau} + d\,.
	\end{equation*}

    The ENs have superior computing capabilities compared to the users. In particular, we assume that the users need $\delta$ normalized time units to perform $r$ multiplications and $r-1$ additions.
	
	\subsection{Communication}
	\label{sec:Comm}
	
	The users have to upload their data to the ENs as well as download the results of the computations from the ENs. We denote by $\gamma$ the normalized time it takes for both upload and download to unicast a symbol $\alpha \in \textrm{GF}{(q)}^u$ (i.e., an element from $\textrm{GF}{(q)}$ for each user). Consequently, the normalized time incurred by all users uploading their data (i.e., $u$ vectors in GF$(q)^r$) to a single EN is
	\begin{equation*}
	    {\mathsf{L}}^\mathsf{comm,up} = \gamma r\,.
	\end{equation*}
	
	For the upload of the private data $\{\x_i\}$ from the $u$ users to the $e$ ENs, we assume that transmission occurs sequentially,  i.e., we consider time-division multiple access, whereas in the download the ENs can transmit simultaneously to multiple users %
	by utilizing joint beamforming based on zero-forcing precoding \cite{Tao,Kuikui,Osvaldo,TaoStudy,Zhang,Naderializadeh}. More precisely, a symbol available at $\rho$ ENs can be transmitted simultaneously to $\min\{\rho,u\}$ users with a normalized communication latency of \(\gamma/\min\{\rho,u\}\) in the high signal-to-noise (SNR) regime. The normalized communication latency in the download (in the high SNR regime) incurred by transmitting $v$ symbols $\alpha_1,\ldots,\alpha_{v}$ in GF$(q)^u$, one element from GF$(q)$ for each user, where symbol $\alpha_i$ is available at $\rho_i$ ENs, is
	\begin{equation*}
		{\mathsf{L}}^\mathsf{comm,down} = \gamma \sum_{i=1}^{v} \frac{1}{\min\{\rho_i,u\}}\,.
	\end{equation*}

The communication latency is then 
\begin{displaymath}
\mathsf{L}^\mathsf{comm}=\mathsf{L}^\mathsf{comm,up}+\mathsf{L}^\mathsf{comm,down}\,.
\end{displaymath}

	\subsection{Privacy and Problem Formulation}
	
    The ENs may not be trustworthy or may be compromised. Further, the compromised ENs may collaborate to infer the data of the users. %
    In this paper, we assume that up to $z$ ENs may be compromised and may collude.
    Our goal is to perform the inference problem over $e$ ENs privately (so that the compromised ENs gain no information in an information-theoretic sense about the private data) while minimizing the overall latency,
        	\begin{align*}
		{\mathsf{L}} &= {\mathsf{L}}^\mathsf{comp} + {\mathsf{L}}^\mathsf{comm} + {\mathsf{L}}^\mathsf{dec}\,,
	\end{align*}
encompassing the  computation and communication latency, as well as the latency incurred by the decoding operation, denoted by $\mathsf{L}^\mathsf{dec}$ and discussed in Section~\ref{sec:decoding}.

%% file: PrivateDistributedLinearInference.tex
	\section{Private Distributed Linear Inference}\label{Sec: PrDiLiIn}
	
	In this section, we present a privacy-preserving coded scheme  that allows $u$ users to perform the linear inference $\{\w\x_i\}$ over $e$ ENs without revealing any information to any subset of $z$ colluding   ENs. A distinguishing feature of the proposed scheme is that, unlike the (nonprivate) scheme in \cite{Osvaldo}, which yields straggler mitigation by introducing redundancy on matrix $\w$, it introduces redundancy on the users' data, by means of secret sharing according to \cite{Shamir}, which allows to achieve straggler mitigation while guaranteeing user data privacy simultaneously. %
	
	\subsection{Secret Sharing}
	
	We consider the SSS scheme to yield privacy. 
	An $(n,k)$ SSS scheme divides a secret into $n$ pieces, referred to as \emph{shares}, such that any $k$ or more shares are sufficient to recover the data, while less than $k$ shares do not reveal any information about data.
    	
    	The proposed scheme is as follows.
    	Each user $\mathsf{u}_i$ uses an \((n, k)\) SSS scheme to compute \(n\) shares of its private data  \(\x_i = (x_{i,1},\ldots,x_{i,r})^\top\). In particular, user $\mathsf{u}_i$  encodes each data entry \(x_{i,l}\) along with $k-1$ i.i.d. uniform random symbols $r_{i,l}^{(1)},\ldots,r_{i,l}^{(k-1)}$ from $\text{GF}(q)$ using a nonsystematic \((n,k)\) Reed-Solomon (RS) code over GF$(q)$ to obtain \(n\) coded symbols  $s_{i,l}^{(1)},\ldots ,s_{i,l}^{(n)}$. Let $\{\bm r_i^{(\kappa)}=(r^{(\kappa)}_{i,1},\ldots,r^{(\kappa)}_{i,r})^\top \mid\kappa \in [k-1]\}$  be the set of vectors of uniform random symbols used by user $\mathsf{u}_i$. For each $h\in[n]$, the $h$-th share of user $\mathsf{u}_i$ is 
    	\begin{align*}
    		\s^{(h)}_i=\left(\begin{matrix}
    			s^{(h)}_{i,1},  \ldots, s^{(h)}_{i,r}\\
    		\end{matrix}\right)^\top.
    	\end{align*}
    	We collect the $h$-th share of all users in the matrix
    	\begin{align}
    	\label{eq:matrix_of_shares}
    		\bm S^{(h)}=\left(\begin{matrix}
    			\s^{(h)}_1, \s^{(h)}_2, \ldots, \s^{(h)}_{u}
    		\end{matrix}\right)\in\text{GF}{(q)}^{r\times u}\,.
    	\end{align}

    	The following theorem proves that the  linear inference operations \(\{\by_i=\w\x_i\}\) can be computed for all users  from a given set of computations based on the matrices of shares $\boldsymbol{S}^{(1)},\ldots, \boldsymbol{S}^{(n)}$,
    	while  providing  privacy against up to  $k-1$ colluding ENs\textemdash which collectively have access to up to $k-1$ distinct matrices of shares.

    	\begin{theorem}\label{Th: RecoverSecretComputation}
    		Consider \(u\) users with their respective private data $\x_i\in\textup{GF}(q)^r$, $i\in[u]$. Use an \((n,k)\) SSS scheme on each \(\x_i\) to obtain the matrices of shares \(\boldsymbol{S}^{(1)},\ldots, \boldsymbol{S}^{(n)}\) in \eqref{eq:matrix_of_shares}. %
    		Let \(\w\in\textup{GF}{(q)}^{m\times r}\) be a public matrix and \(\mathcal I\subseteq[n]\) a set of indices with cardinality \(|\mathcal I|=k\). Then, the set of computations \(\{\w\bm S^{(h)}\mid h\in\mathcal I\}\) allows to recover the computations \(\{\w\x_i\}\) of all users. Moreover, for any set $\mathcal{J}\subseteq[n]$ with $|\mathcal{J}| < k$, \(\{\w\bm S^{(h)}\mid h\in\mathcal J\}\) reveals no information about  \(\{\w\x_i\}\).
    	\end{theorem}
    	The proof is given in Appendix \ref{Ap: RecoverSecretComputation}. 
    	The following corollary gives a sufficient condition to recover the private computations \(\{\w\x_i\}\).
    	\begin{corollary}[Sufficient recovery condition]\label{Cor: Recovery}
    		Consider an edge computing scenario where the public matrix \(\w\) is partitioned row-wise into \(b\) disjoint submatrices \(\w_\ell\in\textup{GF}{(q)}^{\frac{m}{b}\times r}\), $\ell\in[b]$, and the private data is \(\{\x_i \}\). Then, the private computations \(\{\w\x_i \}\) can be recovered from the computations in the sets 
    		\begin{align} 
    		\label{Eq: RecoveryCondition}
    			\set{S}_\ell\triangleq\{\w_\ell\bm S^{(h)}\mid h\in\mathcal I\},\; \ell\in[b]\, ,
    		\end{align}
    		for any fixed set $\mathcal{I}\subseteq[n]$ with cardinality $|\mathcal{I}| = k$.  
    	\end{corollary}
    	\begin{IEEEproof}
    		From \cref{Th: RecoverSecretComputation}, for a given \(\ell\in[b]\),  the computations in the set $\{\w_\ell\x_i\}$ can be recovered from the computations in the set $\set{S}_\ell$. Then, we obtain
    		\[\w\x_i=\left(\begin{matrix}
    			(\w_1\x_i)^\top, (\w_2\x_i)^\top, \ldots, (\w_{b}\x_i)^\top
    		\end{matrix}\right)^\top,~\forall i\in[u]\, .\]
    	\end{IEEEproof}
    	
    	Given the SSS scheme, the proposed scheme can be reduced to two combinatorial problems: the assignment of submatrices $\{\w_\ell\}$ to the ENs such that no EN stores more than a fraction $\mu$ of $\w$, and the assignment of matrices of shares $\{\bm S^{(h)}\}$ to the ENs such that no $z$ colluding ENs gain any information about the data $\{\x_i\}$. We require that the combination of the assignments guarantees the users to obtain the compuations in \cref{Eq: RecoveryCondition}, such that the users have access to sufficient data to recover $\{\w\x_i\}$. In the following two subsections, we describe the assignment of $\{\w_\ell\}$ and $\{\bm S^{(h)}\}$ to the ENs.
	  
	\subsection{Assignment of \(\w\) to the Edge Nodes}\label{Sec: Wassignment}

	    To create joint beamforming opportunities in the download, we allow for replications of the same $\w_\ell$ across different ENs. Submatrices are assigned to the ENs as follows.	In order to satisfy the storage constraint, i.e., no EN can store more than a fraction $\mu$ of $\w$,		we select \(p\in\mathbb{N}\) such that \(p/e \leq \mu\) and partition \(\w\) row-wise into \(e\) submatrices as
		\begin{align*}
			\w=\left(\begin{matrix}
				\w_1^\top, \w_2^\top, \ldots, \w_{e}^\top
			\end{matrix}\right)^\top.	
		\end{align*}
		We then assign \(p\) submatrices to each of the \(e\) ENs. To this scope, we define a  matrix of indices $\bm I_{\mathsf{w}}$, of dimensions $p\times e$, which prescribes the assignment of submatrices to the ENs. The assignment has the following combinatorial structure. Consider a cyclic permutation group of order \(e\) with generator \(\pi\). We construct $\bm I_{\mathsf{w}}$ as 
		\begin{align}
			\label{Eq: Iw}
			\bm{I}_{\mathsf w} &= \left(\begin{matrix}
			\pi^0(1) & \pi^0(2) & \cdots & \pi^0(e)\\
			\pi^1(1) & \pi^1(2) & \cdots & \pi^1(e)\\
			\vdots & \vdots & \ddots & \vdots\\
			\pi^{p-1}(1) & \pi^{p-1}(2) & \cdots & \pi^{p-1}(e)\\
			\end{matrix}\right)
		\end{align}
		and define the set of indices
		\begin{equation}
			\label{Eq:setIw}
			\set{I}_j^{\mathsf w}=\{\pi^{0}(j),\ldots,\pi^{p-1}(j)\}
		\end{equation}
		for $j\in[e]$ as the set containing the entries in column \(j\) of \(\bm I_{\mathsf w}\). Then, we assign the submatrices \(\{\w_\ell\mid \ell\in\set{I}_j^{\mathsf w}\}\) to EN $\mathsf{e}_j$, i.e., $\mathsf{e}_j$ is assigned the submatrices $\{\w_\ell\}$ with indices $\ell$ in the $j$-th column of $\bm I_{\mathsf{w}}$. For example, if \(\pi=(1\;e\; e-1\; \cdots\; 2)\), we have 
		\begin{align*}
			\bm{I}_{\mathsf w}=\left(\begin{matrix}
			1 & 2 & \cdots & e\\
			e & 1 & \cdots & e-1\\
			\vdots & \vdots & \ddots & \vdots\\
			e-p+2 & e-p+3 & \cdots & e-p+1
			\end{matrix}\right)\, ,
		\end{align*}
		and EN $\mathsf{e}_2$ stores $\w_2, \w_{1}, \w_{e},\ldots,\w_{e-p+3}$. 
		
		This assignment of submatrices to ENs bears some resemblance with fractional repetition (FR) codes \cite{ElRouayheb}. FR codes were proposed in the context of distributed storage systems and yield the property that any $\zeta$ storage nodes have access to at least $\psi$ distinct symbols/packets of a $\psi$-dimensional MDS code such that users can recover the data by decoding the MDS code after contacting $\zeta$ storage nodes. By guaranteeing that all pairs of storage nodes share exactly $\theta$ packets (utilizing Steiner systems such as the Fano plane), the authors can derive lower bounds on the number of distinct packets across $\zeta$ storage nodes. From this lower bound, the above-mentioned property (i.e., that any $\zeta$ storage nodes have access to at least $\psi$ distinct symbols/packets) follows. In contrast, our goal is to achieve significant replication of submatrices $\bm W_\ell$ at the ENs, which we achieve by a cyclic structure. We do not have the requirement that any two ENs share exactly $\theta$ packets. Furthermore, one of our proposed schemes (introduced in Section~\ref{sec:SchemeW}) allows for irregular repetition of   packets across ENs, while an essential requirement of FR codes is that packets are repeated the same amount of times across nodes. To summarize, both our assignment of submatrices and FR codes are combinatorial designs, but serve different purposes. Notably, our assignment is much less structured than FR codes.
		
		The ENs process the assigned submatrices of $\w$ in the same order as their indices appear in the rows of $\bm{I}_{\mathsf{w}}$. We define $\phi_j^\mathsf{w}(\ell')$ for $\ell'\in [p]$ to be the map from $\ell'$ to the index of the $\ell'$-th assigned submatrix of EN $\mathsf{e}_j$.

	\subsection{Assignment of Shares to the Edge Nodes}
	\label{sec:assignment}

    On the basis of the assignment of the submatrices of $\w$, to guarantee privacy, we now have to define the assignment of matrices of shares such that no $z$ colluding ENs have access to $k$ or more distinct matrices of shares, while the users should be guaranteed to obtain the computations in \cref{Eq: RecoveryCondition}. Here, we restrict the number of shares $n$ to be at most equal to the number of ENs, i.e., we require $n\leq e$.
    As with the submatrices of $\w$, we  allow replicating shares across ENs to exploit joint beamforming opportunities in the download. However, this may lead to multiple shares being assigned to a single EN, which presents difficulties in the design of a private scheme, because having multiple shares available at a single EN results in a privacy level $z$ lower than that of the SSS scheme ($k$). For example, if all ENs have access to two matrices of shares, the scheme only provides privacy against any $z=\lfloor (k-1)/2 \rfloor$ colluding ENs.
    
    Alike to $\bm I_{\mathsf{w}}$, let $\bm I_{\mathsf{s}}$  be the index matrix that prescribes the assignment of matrices of shares to the ENs\textemdash the users upload their shares to the ENs according to $\bm I_{\mathsf{s}}$. The assignment has the following structure. Given the generator \(\pi\) used to assign the submatrices of \(\w\) to the ENs, we construct the \((\beta+1)\times e\) index matrix $\bm I_{\mathsf{s}}$ as 
	\begin{align}
		\label{Eq: Is}
		\bm I_{\mathsf s}=\Scale[0.972]{\left(\begin{matrix}
			\pi^{0}(1) & \pi^{0}(2) & \cdots & \pi^{0}(e)\\
			\pi^{e-p}(1) & \pi^{e-p}(2) & \cdots & \pi^{e-p}(e)\\
			\vdots & \vdots & \ddots & \vdots\\
			\pi^{\beta(e-p)}(1) & \pi^{\beta(e-p)}(2) & \cdots & \pi^{\beta(e-p)}(e)
		\end{matrix}\right)}\, ,
	\end{align}
	where \(\beta=\left\lceil e/p\right\rceil-1\). Define the set of indices
	\begin{equation}
		\label{Eq:setIs}
		\mathcal{I}_j^{\mathsf{s}}=\{\pi^{0}(j),\ldots,\pi^{\beta(e-p)}(j)\}\backslash\{n+1, n+2,\ldots,e\}
	\end{equation}
	as the subset of entries in column \(j\) of \(\bm I_{\mathsf s}\) that are in $[n]$. We have 
	\begin{align}
	\label{eq:cardIsj}
	    |\mathcal{I}_j^{\mathsf{s}}|  = \left\lceil \lceil e/p\rceil \cdot n/e\right\rceil \triangleq a\,, 
	\end{align}
	as we keep only a fraction $\left\lceil n/e \right\rceil$ of the shares corresponding to the $\beta + 1 = \left\lceil e/p\right\rceil$ used permutations in \(\bm I_{\mathsf s}\). Then, user $\mathsf{u}_i$ transmits the shares $\{\bm s_i^{(h)}\mid h\in\mathcal{I}_j^{\mathsf{s}}\}$ to EN \(\mathsf{e}_j\), i.e., EN  $\mathsf{e}_j$ is assigned $a$ matrices of shares $\{\bm S^{(h)}\}$ with indices $h$ in the $j$-th column of $\bm I_{\mathsf s}$ that are in $[n]$. Consequently, $z$ colluding ENs have access to $az$ possibly distinct matrices of shares. To guarantee user data privacy against any subset of $z$ colluding ENs, we have to impose the constraint $k\geq az+1$.
	
	Similar to the submatrices of $\w$, the shares are processed by the ENs in the same order as their indices appear in the rows of $\bm{I}_{\mathsf{s}}$. We define $\phi_j^\mathsf{s}(h')$ for $h'\in [a]$ to be the map from $h'$ to the index of the $h'$-th assigned matrix of shares of EN $\mathsf{e}_j$. %
	For all $\ell'\in[p]$, EN $\mathsf{e}_j$ computes $\w_{\phi_j^\mathsf{w}(\ell')}\bm S^{(\phi_j^\mathsf{s}(h'))}$ before moving on to the next matrix of shares $\bm S^{(\phi_j^\mathsf{s}(h'+1))}$.
	The following theorem shows that the combined assignment of submatrices and shares to the ENs allow each user $\mathsf{u}_i$ to obtain its desired result $\w\x_i$ while preserving privacy against up to $z$ colluding ENs.
	
	\begin{theorem}\label{Th: recover}
	Consider an edge computing network consisting of \(u\) users and \(e\) ENs, each with a storage capacity corresponding to a fraction  \(\mu\), $0 < \mu \leq 1$, of $\w$, and an \((n,k\ge az+1)\) SSS scheme, with \(n\leq e\) and $a$ given in \eqref{eq:cardIsj}. For \(j\in[e]\), EN \(\mathsf{e}_j\) stores the submatrices of \(\w\) from the set \(\{\w_\ell\mid \ell\in\set{I}_j^{\mathsf w}\}\) with $\set{I}_j^{\mathsf w}$ defined in \cref{Eq:setIw}. Furthermore, it receives the matrices of shares from the set \(\{\bm S^{(h)}\mid h\in\mathcal{I}_j^{\mathsf{s}}\}\) with $\mathcal{I}_j^{\mathsf{s}}$ defined in \cref{Eq:setIs}, and computes and returns the set \(\{\w_\ell\bm S^{(h)} \mid \ell\in\set{I}_j^{\mathsf w},  h\in\mathcal{I}_j^{\mathsf{s}} \}\) to the users. Then, all users can recover their desired computations $\{\w \bm x_i \}$ and the scheme preserves privacy against any set of $z$ colluding ENs.
	\end{theorem}
	The proof of \cref{Th: recover} is given in Appendix \ref{App: recover}. We provide a sense of the proof with the following example.
	\begin{example}
		Consider \(e=n=5\), \(p=3\), and \(\pi=(1\;4\;2\;5\;3)\), the generator of a cyclic permutation group of order \(5\). From \cref{Eq: Iw,Eq: Is}, we have
		\begin{align*}
			\bm I_{\mathsf w}&=\left(\begin{matrix}
				1 & 2 & 3 & 4 & 5\\
				4 & 5 & 1 & 2 & 3\\
				2 & 3 & 4 & 5 & 1
			\end{matrix}\right) \text{ and }
			\bm I_{\mathsf s}&=\left(\begin{matrix}
				1 & 2 & 3 & 4 & 5\\
				2 & 3 & 4 & 5 & 1
			\end{matrix}\right)\,.
		\end{align*}
		We focus on the matrix  of shares $\bm S^{(1)}$. It is assigned to EN $\mathsf{e}_1$ and gets multiplied with the submatrices of $\bm W$ indexed by the elements in the set
		\begin{align*}
		   \set{I}_1^{\mathsf w} = \{\pi^0(1), \pi(1), \pi^2(1)\}=\{1,4,2\}\,.
		\end{align*}
		Note that the set $\set{I}_1^{\mathsf w}$ contains three recursively $\pi$-permuted integers of $1$  ($\pi^0(1)$, $\pi^1(1)$, and $\pi^2(1)$). Now, consider EN $\mathsf{e}_5$, which is also assigned the matrix  of shares $\bm S^{(1)}$. We have %
		\begin{align*}
			\set{I}_5^{\mathsf w} = \{\pi^0(5), \pi(5), \pi^2(5)\}=\{5,3,1\}\,.
		\end{align*}
		Notice that \(\pi^0(5) = \pi^3(1)=5\) is the fourth (including $\pi^0$) recursively $\pi$-permuted integer of $1$. %
		Hence, the set \(\set{I}_1^{\mathsf w}\cup\set{I}_5^{\mathsf w}\) contains in total six recursively $\pi$-permuted integers of $1$,  %
		which is sufficient to give the set \([5]\), since the group generated by $\pi$ is transitive. In a similar way, it can be shown that the same property holds for all other matrices of shares. Each matrix of shares is multiplied with all submatrices of $\bm W$, and the sets in \cref{Eq: RecoveryCondition} are obtained.
	\end{example}

%% file: CommunicationAndComputationScheduling.tex
	\section{Communication and Computation Scheduling, and Private Coding Scheme Optimization}\label{Sec: CoCoSh}
	
	In this section, we describe the scheduling of the proposed scheme. This encompasses the upload of the shares to the ENs, the order of the computations performed at the ENs, the download of a sufficient subset of $\{\w_\ell\boldsymbol{S}^{(h)} \mid \ell \in \mathcal{I}_j^{\mathsf{w}}, h\in\mathcal{I}_j^{\mathsf{s}},j \in [e]\}$, and the decoding of this subset such that each user $\mathsf{u}_i$ obtains the desired result $\by_i=\w\x_i$. In the following, we refer to a product $\w_\ell\boldsymbol{S}^{(h)}$ as an intermediate result (IR).
	
	\subsection{Upload and Computation}
	
\begin{figure}
		\includegraphics[width=\columnwidth]{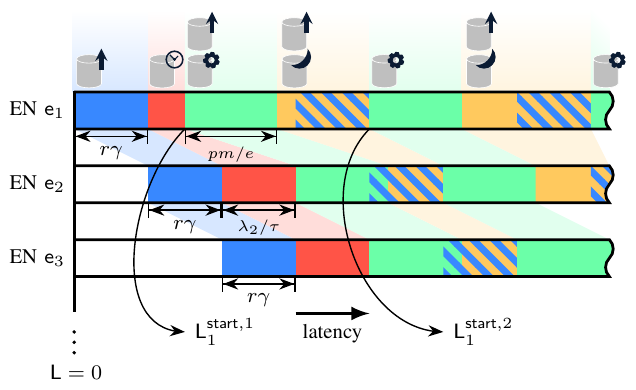}
		\vspace{-3ex}
		\caption{Scheduling of the upload and computing phases. For each EN, the upload normalized times $r\gamma$ are shown in blue, the random setup times in red, the times $pm/e$ to compute $p$ IRs in green, and  possible idle times in yellow.}
		\label{fig:schedule}
	\end{figure}

	Our scheme starts with the users uploading their shares to the ENs. As $\w$ stays constant over a long period of time, we assume that it  can be stored at the ENs prior to the beginning of the online phase. The users start by sequentially unicasting their shares to the $e$ ENs. Note that, unlike in the nonprivate scheme in \cite{Osvaldo}, the users cannot broadcast their data in the clear\textemdash to attain privacy, it needs to be ensured that any $z$  potentially compromised ENs do not gain access to more than $k-1$ distinct shares of the users' private data. Recall that transmission of one element of $\text{GF}(q)$ from each user takes  $\gamma$ normalized time units (see Section~\ref{sec:Comm}). Consequently, it takes $\gamma r$ time units until an EN receives a matrix of shares $\boldsymbol{S}^{(h)}$. The upload schedule is depicted in blue in \cref{fig:schedule}. The users first upload their first matrix of shares to EN $\mathsf{e}_1$ and continue with $\mathsf{e}_2,\mathsf{e}_3,\ldots$ sequentially until each EN has received its first matrix of shares. The users then transmit their second matrix of shares to the $e$ ENs, starting with $\mathsf{e}_1$. This continues until each EN  has received  $a$ matrices of shares; EN $\mathsf{e}_j$ receives $\{\boldsymbol{S}^{(h)} \mid h\in\mathcal{I}_j^{\mathsf{s}}\}$. Hence, EN $\mathsf{e}_j$ receives its $h'$-th matrix of shares, $\bm S^{(\phi^{\mathsf{s}}_j(h'))}$, at normalized time
	\begin{equation*}
		{\mathsf{L}}^{\mathsf{up},h'}_j =  \gamma r(e(h'-1) + j)\,,
	\end{equation*}
	and the total normalized upload latency of the private scheme becomes
	\begin{align*}
	{\mathsf{L}}^\mathsf{up}_\mathsf{P} = \gamma \cdot r\cdot e\cdot a\,.
	\end{align*}
    
    The computation phase at an EN starts as soon as the EN receives the first matrix of shares from the users. Recall from Section~\ref{sec:RunTime} that 
    the random setup time  for EN $\mathsf{e}_j$ is $\lambda_j$, i.e., EN $\mathsf{e}_j$ starts the computation $\lambda_j/\tau$ normalized time units after receiving its first matrix of shares. The setup times are illustrated in red in \cref{fig:schedule}. In total, $p$ IRs of the form $\w_\ell\bm S^{(h)}$ have to be computed for each assigned matrix of shares $\bm S^{(h)}$ by EN $\mathsf{e}_j$, $j\in[e]$, where  $\ell \in  \set{I}_j^{\mathsf w}$ and $h \in  \set{I}_j^{\mathsf s}$. This incurs a normalized latency of $p\cdot m/e$, because each $\w_\ell$ has $m/e$ rows, and hence the ENs compute $u\cdot m/e$ inner products for each of the $p$ IRs.
    
    It can happen that an EN has not received the next matrix of shares when it finished the computation on the current matrix of shares. In this case, the EN remains idle until the users upload the next matrix of shares. We depict this in yellow in \cref{fig:schedule}. For $h' \in [a]$, the normalized time at which EN $\mathsf{e}_j$ starts to compute on the $h'$-th assigned matrix of shares, i.e., on $\bm S^{(\phi^{\mathsf{s}}_j(h'))}$, is 
	\begin{equation*}
		\mathsf{L}^{\mathsf{start},h'}_j = \max\left \{\mathsf{L}^{\mathsf{start},h'-1}_j + p\frac{m}{e}~,~\mathsf{L}^{\mathsf{up},h'}_j\right\},\; \text{for $h' > 1$}\,,
	\end{equation*}
	and
	\begin{equation*}
		\mathsf{L}^{\mathsf{start},1}_j = \frac{\lambda_j}{\tau} + \mathsf{L}^{\mathsf{up},1}_j\,.
	\end{equation*}
	
	The computation phase continues at least until the computations in \cref{Eq: RecoveryCondition} are obtained, i.e.,  until there are at least $k$ distinct IRs of the form $\w_\ell \bm S^{(h)}$  for each $\ell \in [e]$. This ensures that  user $\mathsf{u}_i$ can recover  \(\w\x_i\). We remark that it can be beneficial to continue computing products to reduce the communication latency in the download phase, as we discuss next.

	\subsection{Download}
	
	For the download, we exploit zero-forcing precoding  to  serve multiple users simultaneously and hence reduce the communication latency. An IR $\w_\ell\boldsymbol{S}^{(h)}$ that is available at $\rho_{\ell,h}$ ENs incurs a normalized communication latency of $(m/e)\cdot\gamma /\min\{\rho_{\ell,h},u\}$ (see Section~\ref{sec:Comm}). Consequently, a high multiplicity of an IR reduces its corresponding communication latency. However, a high multiplicity implies that the same IR has to be computed multiple times at different ENs, thereby increasing the computation latency. There is therefore a tradeoff between communication latency and computation latency, which  can be optimized to reduce the overall latency. Assume the optimum is reached after EN $\mathsf{e}_{j^*}$ has computed the IR $\w_{\phi^\mathsf{w}_{j^*}(\ell^*)}\boldsymbol{S}^{(\phi^\mathsf{s}_{j^*}(h^*))}$. This gives a normalized computation latency of
	
	\begin{equation*}
		{\mathsf{L}}^\mathsf{comp} = {\mathsf{L}}^{\mathsf{start},h^*}_{j^*} + \ell^* \frac{m}{e}\,.
	\end{equation*}
	
	Subsequently, the ENs jointly transmit a subset of the computed IRs $\{\w_\ell\boldsymbol{S}^{(h)}\}$ to multiple users simultaneously in descending order of their multiplicity $\rho_{\ell,h}$ until enough results are available to the users such that the sufficient recovery condition in \cref{Cor: Recovery} is met. More precisely, for each $\w_\ell$ the ENs send the $k$ IRs with  highest multiplicity to the users, thereby ensuring that each user $\mathsf{u}_i$ can recover the desired result $\w\x_i$. For a fixed $\ell$, let 
	\begin{align*}
	\mathcal{H}_\ell^\mathsf{max} = \arg\max_{\mathcal{A}\subseteq[n],|\mathcal{A}| = k} \sum_{h\in\mathcal{A}} \rho_{\ell,h}
	\end{align*}
	be the set of indices $h$ of the $k$ largest $\rho_{\ell,h}$. Then, the aforementioned download strategy results in a normalized communication latency of
		\begin{equation*}
		{\mathsf{L}}^\mathsf{comm} = \gamma\frac{m}{e} \sum_{\ell = 1}^{e}\sum_{h\in \mathcal{H}_\ell^\mathsf{max}} \frac{1}{\min\{\rho_{\ell,h},u\}}\,.
	\end{equation*}

	\subsection{Decoding Latency}
	\label{sec:decoding}
	After the users have downloaded a sufficient number of IRs ($k$ IRs for each $\w_\ell$), the users need to decode the SSS scheme to obtain their desired results $\{\bm y_i=\w\x_i\}$. Decoding  the SSS scheme means decoding the corresponding RS code. Here, we assume decoding via the Berlekamp-Massey algorithm, which, for an $(n,k)$ RS code, entails $n(n - k)$ multiplications and $n(n - k - 1)$ additions \cite{Garrammone}, plus an additional discrete Fourier transformation that involves $n/2(\lceil\log_2(n)\rceil-1)$ multiplications and $n\lceil\log_2(n)\rceil$ additions \cite[Eq. (8)]{Yavne}.
	We assume that it takes the same time to perform one addition and one multiplication, i.e., both operations take the same amount of clock cycles. This assumption is reasonable, as both operations can be performed using either a look-up table or, in case $q$ is a prime, integer arithmetic in the arithmetic and logic units of the user devices' processors. We make this assumption because it significantly simplifies  the analysis. Recall that a user requires $\delta$ normalized time units to compute an inner product in $\mathrm{GF}{(q)}^r$, which comprises $r$ multiplications and $r-1$ additions in $\mathrm{GF}{(q)}$. The  latency of performing an addition or a multiplication is hence $\delta/(2r-1)$. With this, the decoding latency for each user can be written in closed-form as
	 \begin{equation*}
	    {\mathsf{L}}^\mathsf{dec} = \frac{\delta}{2r-1} m n\left( 2(n-k)+\frac{3}{2}\lceil\log_2(n)\rceil-\frac{3}{2}\right)\,,
	\end{equation*}
	since the users have to perform 
	\begin{equation*}
	    \frac{m}{e} \cdot e \cdot n\left( 2(n-k)+\frac{3}{2}\lceil\log_2(n)\rceil-\frac{3}{2}\right)
	\end{equation*} operations in $\mathrm{GF}{(q)}$ (one RS decoding per row) for each of the $e$ matrices $\{\w_\ell\}$ while needing  $\delta/(2r-1)$ normalized times units per operation.

	The overall normalized latency becomes
	\begin{align}
	\label{eq:overall_latency}
		{\mathsf{L}} &= {\mathsf{L}}^\mathsf{comp} + {\mathsf{L}}^\mathsf{comm} + {\mathsf{L}}^\mathsf{dec}\nonumber\\
		&= {\mathsf{L}}^{\mathsf{start},h^*}_{j^*} + \ell^* \frac{m}{e}+ \gamma\frac{m}{e} \sum_{\ell = 1}^{e}\sum_{h\in \mathcal{H}_\ell^\mathsf{max}} \frac{1}{\min\{\rho_{\ell,h},u\}} \nonumber\\
		&~~~~+\frac{\delta}{2r-1} m n\left( 2(n-k)+\frac{3}{2}\lceil\log_2(n)\rceil-\frac{3}{2}\right)\,.
	\end{align}

\subsection{Private Coding Scheme Optimization}	
	
	The system design includes the SSS code, which we denote by $\mathscr{C}_{\mathsf{SSS}}$, the assignment matrices $\bm I_{\mathsf{w}}$ and $\bm I_{\mathsf{s}}$, the number of ENs over which the users offload the linear inference operation, $e\le e_{\mathsf{max}}$, and the privacy level $z$. To construct matrices $\bm I_{\mathsf{w}}$ and $\bm I_{\mathsf{s}}$, we require a permutation group generator $\pi$ and the parameter $p$.
	Further, to determine a good stopping point for the computation phase, we introduce the parameter $t$, defined as the  number of (not necessarily distinct) IRs for each $\w_\ell$ computed across all ENs to wait for before the download phase starts. %
	Note that $t$ should be such that the ENs have collected enough distinct IRs so that the users can decode to recover $\{\by_i=\w_i\x_i\}$. However, it might be useful to collect more IRs than the minimum necessary to reduce the communication latency. As soon as there are $t$ (not necessarily distinct) IRs computed across all ENs for each submatrix of $\w$, the ENs stop the computation and begin the download phase.

	We refer to the tuple $(\mathscr{C}_{\mathsf{SSS}},e,\pi,p,t,z)$ as the \emph{private coding scheme}. The goal is then to optimize the private coding scheme, i.e., the above-mentioned tuple,
in order to minimize $\mathsf{L}$ in \eqref{eq:overall_latency} for a given privacy level $z$.

    Note that $e\le \emax$ (it may be beneficial to contact less ENs than the ones available). Furthermore,  even for the lowest level of privacy, $z=1$, the users need to contact at least $2$ ENs, i.e., $2\leq e\leq e_{\max}$. %
	Additionally, $1\leq p \leq \lfloor \mu e \rfloor$; each EN needs  to be assigned at least one partition of $\w$ and it may be beneficial that the ENs do not utilize their whole storage capacity, because storing fewer than $\lfloor \mu e \rfloor$ submatrices of $\w$ leads to the ENs performing computations on the later shares sooner.
	Lastly, there are some constraints on the parameters $n$ and $k$ of the SSS code $\code_{\mathsf{SSS}}$. From the SSS code, it follows that $n\ge k$, whereas from the combinatorial design $n\le e$. The value of $k$ depends on the desired privacy level $z$ and the number of matrices of shares each EN has access to, $a$ (which follows from $e$, $p$, and $n$, see \eqref{eq:cardIsj}). In the worst case, $z$ ENs have access to $az$ distinct shares. Therefore, we need $k\geq az +1$ to  ensure privacy. Consequently, we get $az+1\leq n \leq e$. Note that there is no reason to select $k > az +1$ as this leads to reduced straggler mitigation and increased computational load.

%% file: Variants.tex
\section{Variants}\label{Sec: Variants}

    In this section, we  introduce two variants to the private scheme proposed in Sections~\ref{Sec: PrDiLiIn} and~\ref{Sec: CoCoSh}. First, we notice that we can reduce the overall latency by starting the download as soon as the upload phase is completed, i.e., the download phase and the computation phase can be performed  simultaneously. We propose to use a priority queue that determines the order in which computed IRs should be downloaded. Secondly, we introduce an additional layer of coding to the scheme by encoding the network-side matrix $\w$ prior to storing it over the ENs. %
    We also relax some of the constraints on the system parameters.
    
    \subsection{Priority Queue}

 Instead of waiting for the computation phase to finish,  IRs can be downloaded as soon as they are available and the channel is idle, i.e., when the upload phase is completed and no other IR is being downloaded. To determine which IR to send, we equip the ENs with a shared priority queue in which the pairs of indices that identify the IRs, $(\ell,h)$ (where $\ell$ identifies the partition of $\w$, $\w_\ell$, and $h$ the matrix of shares, $\bm S^{(h)}$), are queued. A priority queue is a data structure  in which each element has an associated priority. Elements with high priority will leave the queue before elements with low priority. Particularly, we consider a priority queue in which the priority is given by the multiplicity of an IR.  After an EN has finished the computation of an IR, it either adds the corresponding pair of indices $(\ell,h)$ to the queue or increments its multiplicity (priority) if it already exists in the queue. Anytime the channel is available and there are index pairs left in the queue, the ENs cooperatively send the corresponding IR with the highest priority (i.e., highest multiplicity) to the users. This ensures that at any time the ENs send the IR with the lowest associated communication cost to the users. In contrast to the scheme in Sections~\ref{Sec: PrDiLiIn} and~\ref{Sec: CoCoSh}, there is no optimization needed to determine $t$, as
  the download starts as soon as the upload phase finishes. Hence, the optimization is over $(\mathscr{C}_{\mathsf{SSS}},e,\pi,p,z)$ for a given value of $z$.  Further, the ENs have to keep track of the queue and its complete history. This way, already downloaded IRs do not need to be computed again.

	\subsection{Additional Coding on the Network-Side Matrix $\w$}
	\label{sec:SchemeW}
	
	The straggler resiliency of the scheme proposed in the previous sections can be increased by introducing an additional layer of coding on the matrix $\w$. %
	In particular, we partition $\w$ row-wise into $k'$ submatrices and encode it using an $(n',k')$ RS code, denoted by $\mathscr{C}_{\mathsf{w}}$. We denote by $\bC=\left(\begin{matrix}
				\bC_1^\top, \bC_2^\top, \ldots, \bC_{n'}^\top
			\end{matrix}\right)^\top$ the resulting coded matrix, comprising $n'$ submatrices. The $n'$ coded submatrices of $\bC$ are then assigned to the ENs. Compared to the uncoded case, we relax the condition that the number of submatrices equals the number of ENs $e$.
		For $n'=e$, the same assignment of submatrices to ENs as the one used in Section~\ref{Sec: Wassignment} for the uncoded matrix $\w$ can be used.
	For  $n' \neq e$, however, we need to modify the assignment. For $n'  \geq e$, we simply take a cyclic permutation group $\pi_{\mathsf c}$ of order $n'$ to fill the index matrix $\bm I_{\mathsf c}$ that determines the assignment of submatrices $\{\bC_{\ell}\}$ to ENs (i.e., $\bm I_{\mathsf c}$ is the counterpart of $\bm I_{\mathsf w}$ for the uncoded case and $\pi_{\mathsf c}$ is the counterpart of $\pi$, see Section~\ref{Sec: Wassignment}). Using the same approach for $n' < e$ works, but it leads to a nonuniform distribution of indices in $\bm I_\mathsf{c}$. This would lead to  higher multiplicity for some IRs, which is suboptimal in terms of download latency. Increasing the multiplicity of IRs has diminishing returns; increasing the multiplicity from $1$ to $2$ reduces the communication latency by $50\%$, whereas increasing the multiplicity from $2$ to $3$ yields a decrease of only $33.3\%$. This means that the highest gains are obtained by increasing the multiplicity simultaneously across IRs, i.e., we are interested in obtaining a distribution of indices in $\bm I_\mathsf{c}$ as close as possible to a uniform distribution. To accomplish that, we propose the following index assignment for $n'<e$.
 We start by cyclically filling $\bm{I}_{\mathsf c}$ with indices in $[n']$,
	\begin{equation}
	\label{Eq: Ic}
    	\bm I_{\mathsf c}=\left(\begin{matrix}
			1 & 2 & \cdots & n' & ? & ? & \cdots & ?\\
			? & 1 & 2 & \cdots & n' & ? & \cdots & ?\\
			\vdots & \ddots & \ddots & \ddots & \ddots & \ddots & \ddots & \vdots\\
			? & \cdots & ? & 1 & 2 & \cdots & n' & ?\\
		\end{matrix}\right)\,. 
	\end{equation}
	The left-out entries marked with $?$ are filled such that the distribution of indices in $\bm I_{\mathsf c}$ is as close to uniform as possible while not repeating indices in the same column of $\bm I_{\mathsf c}$, as this assignment does not favor a specific submatrix of $\bC$ and prevents the same submatrix being assigned twice to one EN. Note that \cref{Eq: Ic} is only one example of how $\bm I_{\mathsf{c}}$ can look like. Depending on $e$, $p$, and $n'$ there might be wrap-arounds of indices.
	
	We can also relax the condition that the number of secret shares has to be less than or equal to the number of ENs, i.e., we allow $n> e$, and simply consider a permutation group $\pi_{\mathsf s}$ of order $\max(n,e)$ and construct $\bm I_{\mathsf s}$ as in  \eqref{Eq: Is} (with $\pi=\pi_{\mathsf s}$).%
	
	\begin{remark}
	By allowing $n'\neq e$ and $n > e$, it becomes difficult to prove a similar result as in Corollary~\ref{Cor: Recovery} for the uncoded case on a	sufficient condition on the cardinality of $\mathcal I$ and $\mathcal J$ such that the linear inference can be completed from the IRs $\{\bC_\ell\bm S^{(h)}\mid h\in\mathcal I, \ell \in \mathcal J\}$.
	However, our numerical results reveal that encoding $\w$ and relaxing the constraints $n'= e$ and $n \le e$ allows to reduce the overall latency compared to the scheme in Sections~\ref{Sec: PrDiLiIn} and~\ref{Sec: CoCoSh}.
	\end{remark}

	For each user $\mathsf{u}_i$\textemdash with its private data $\x_i$ and set of random vectors $\{\bm r_i^{(1)},\ldots,\bm r_i^{(k-1)}\}$\textemdash the combination of the $(n,k)$ RS code on $\{\x_i,{\bm r}_i^{(1)},\ldots,{\bm r}_i^{(k-1)}\}$ used in the SSS scheme and the $(n',k')$ RS code on $\{\w_1,\ldots,\w_{k'}\}$ can be seen as an $(n n', k k')$ product code (with nonsystematic component codes) over $\{\w_\ell\x_i|\ell\in[k']\}$ (i.e., the desired inference $\w\x_i$) and $\{\w_\ell\bm r_i^{(\kappa)}|\ell\in[k'],\kappa\in[k-1]\}$. 
	To show this,  we arrange the elements of $\{\bC_\ell\bm s_i^{(h)}\}$ in the $n'\times n$ two-dimensional  array 
	\begin{equation*}
	    \left[
	    \begin{array}{cccc}
	        \bC_1\bm s_i^{(1)} & \bC_1\bm s_i^{(2)} & \cdots & \bC_1\bm s_i^{(n)}  \\
	        \bC_2\bm s_i^{(1)} & \bC_2\bm s_i^{(2)} & \cdots & \bC_2\bm s_i^{(n)}  \\
	        \vdots & \cdots & \ddots & \vdots  \\
	        \bC_{n'}\bm s_i^{(1)} & \bC_{n'}\bm s_i^{(2)} & \cdots & \bC_{n'}\bm s_i^{(n)} 
	    \end{array}
	    \right]\,.
	\end{equation*}
		It is easy to see that  each row of the array is a codeword of an $(n,k)$ code and each column is a codeword of an $(n',k')$ code. %
	More precisely, $(\bm s_i^{(1)},\ldots,\bm s_i^{(n)})$ is the codeword corresponding to the encoding of $(\x_i,\bm r_i^{(1)},\ldots,\bm r_i^{(k-1)})$ via the SSS $(n,k)$ RS code. Since the RS code is linear, $(\bC_\ell\bm s_i^{(1)},\ldots,\bC_\ell\bm s_i^{(n)})$ is also a codeword of an $(n,k)$ RS code, which would result from encoding $(\bC_\ell\x_i,\bC_\ell\bm r_i^{(1)},\ldots,\bC_\ell\bm r_i^{(k-1)})$. Likewise, $(\bC_1,\ldots,\bC_{n'})$ is the codeword corresponding to the encoding of $(\w_1,\ldots,\w_{k'})$ via the $(n',k')$  RS code on $\w$, and $(\bC_1\bm s_i^{(h)},\ldots,\bC_{n'}\bm s_i^{(h)})$ is a codeword of an $(n',k')$ RS code corresponding to the encoding of $(\w_1\bm s_i^{(h)},\ldots,\w_{k'}\bm s_i^{(h)})$. %
	
	The product code structure allows the users to iteratively decode the received results, which provides more flexibility regarding the decodable patterns; there are sets of IRs that allow to complete the linear inference operation by  iterating between row and column decoders, while either component code  would fail to decode on its own. To illustrate the iterative decoding procedure, we provide the following example.
	
	\begin{example}
Consider the SSS $(n,k) = (4,3)$ RS code and an $(n',k') = (3,2)$ RS code on $\w$. Encode $\w$ into a matrix $\bC$ and arrange all $\{\bC_\ell\bm s_i^{(h)}|\ell\in[3],h\in[4]\}$ in an array of dimensions  $3\times 4$ to show the product code structure,
	\begin{equation*}
	    \left[
	    \begin{array}{cccc}
	        \bC_1\bm s_i^{(1)} & \bC_1\bm s_i^{(2)} & \bC_1\bm s_i^{(3)} & \bC_1\bm s_i^{(4)}  \\
	        \bC_2\bm s_i^{(1)} & \bC_2\bm s_i^{(2)} & \bC_2\bm s_i^{(3)} & \bC_2\bm s_i^{(4)}  \\
	        \bC_{3}\bm s_i^{(1)} & \bC_{3}\bm s_i^{(2)} & \bC_3\bm s_i^{(3)} & \bC_{3}\bm s_i^{(4)} 
	    \end{array}
	    \right]\,.
	\end{equation*}
	Each row of the array is a codeword of a $(4,3)$ RS code  and each column is a codeword of a $(3,2)$ RS code. 
	
	Assume that the users have the following IRs available,
	\begin{equation*}
	    \left[
	    \begin{array}{ cccc } 
         \bC_{1}\bm s_i^{(1)} & \bC_{1}\bm s_i^{(2)} &  & \\
         & \bC_{2}\bm s_i^{(2)} & \bC_{2}\bm s_i^{(3)} & \\
         \bC_{3}\bm s_i^{(1)} & & & \bC_{3}\bm s_i^{(4)}\\
        \end{array}
        \right]\,.
	\end{equation*}
	As we can see, there are no $k=3$ IRs available for any $\bC_\ell$. Therefore, the users would not be able to decode any SSS scheme. However, we have $k'=2$ IRs available in the first and second column. The users can then decode the column RS code  for columns one and two %
	 to obtain  $\bC_2\s_i^{(1)}$ and $\bC_3\s_i^{(2)}$,
	\begin{equation*}
	    \left[
	    \begin{array}{ cccc } 
         \bC_{1}\bm s_i^{(1)} & \bC_{1}\bm s_i^{(2)} &  & \\
         {\color{red}\bC_{2}\bm s_i^{(1)}} & \bC_{2}\bm s_i^{(2)} & \bC_{2}\bm s_i^{(3)} & \\
         \bC_{3}\bm s_i^{(1)} & {\color{red}\bC_{3}\bm s_i^{(2)}} &  & \bC_{3}\bm s_i^{(4)}\\
        \end{array}
        \right]\,.
	\end{equation*}
	Now, there are $k=3$ IRs available in the second and third row, hence the users can decode the corresponding row codes to obtain $\bC_2\s_i^{(4)}$ and $\bC_3\s_i^{(3)}$,
	\begin{equation*}
	    \left[
	    \begin{array}{ cccc } 
         \bC_{1}\bm s_i^{(1)} & \bC_{1}\bm s_i^{(2)} &  & \\
         {\color{red}\bC_{2}\bm s_i^{(1)}} & \bC_{2}\bm s_i^{(2)} & \bC_{2}\bm s_i^{(3)} & {\color{mycolor5}\bC_{2}\bm s_i^{(4)}}\\
         \bC_{3}\bm s_i^{(1)} & {\color{red}\bC_{3}\bm s_i^{(2)}} & {\color{mycolor5}\bC_{3}\bm s_i^{(3)}}  & \bC_{3}\bm s_i^{(4)}\\
        \end{array}
        \right]\,.
	\end{equation*}
	Lastly, the users can switch to column decoding again as now there are $k'=2$ IRs available in the third and fourth column, and the whole code array can be recovered,
	\begin{equation*}
	    \left[
	    \begin{array}{ cccc } 
         \bC_{1}\bm s_i^{(1)} & \bC_{1}\bm s_i^{(2)} & {\color{blue} \bC_{1}\bm s_i^{(3)}} & {\color{blue} \bC_{1}\bm s_i^{(4)}} \\
         {\color{red}\bC_{2}\bm s_i^{(1)}} & \bC_{2}\bm s_i^{(2)} & \bC_{2}\bm s_i^{(3)} & {\color{mycolor5}\bC_{2}\bm s_i^{(4)}}\\
         \bC_{3}\bm s_i^{(1)} & {\color{red}\bC_{3}\bm s_i^{(2)}} & {\color{mycolor5}\bC_{3}\bm s_i^{(3)}}  & \bC_{3}\bm s_i^{(4)}\\
        \end{array}
        \right]\,.
	\end{equation*}
	At last, the users are able to recover all IRs and thereby the computations $\{\w\x_i\}$. For this particular example, this would not have been possible without the redundancy on the submatrices of $\w$.
	\end{example}
	
	The private coding scheme with coding over $\w$ and priority queue is defined by the tuple 	$(\mathscr{C}_{\mathsf{SSS}},\mathscr{C}_{\mathsf{w}},e,\pi_{\mathsf c},\pi_{\mathsf s},p,z)$, which should be properly optimized for a given privacy level $z$.

%% file: OptimizationAndNumericalResults.tex
	\section{Numerical Results}\label{Sec: OpNuRe}
	In this section, we compare the performance of the proposed private scheme in Sections~\ref{Sec: PrDiLiIn} and~\ref{Sec: CoCoSh}, and its variants in Section~\ref{Sec: Variants}, 	with that of the nonprivate scheme in \cite{Osvaldo}.
	For convenience,  we will refer to the scheme in Sections~\ref{Sec: PrDiLiIn} and~\ref{Sec: CoCoSh} as Scheme~1, and to the two variants in Section~\ref{Sec: Variants} as Scheme~2 (Scheme~1 augmented with a priority queue) and Scheme~3 (Scheme~2 augmented with coding over $\w$).%

	For all numerical results, the maximum number of ENs is $\emax = 9$, the storage capacity is $\mu = 2/3$, $\w$ has dimensions $600\times 50$, the computation time is $\tau = 0.0005$, and the straggling parameter is $\eta = 0.5$. Lastly, we assume that the users are $\delta = 3$ times slower than the ENs. Note that due to the normalization by $\tau$, the number of users is inconsequential on the normalized overall latency $\mathsf{L}$ as long as $u \geq \max_{\ell,h}~ \rho_{\ell,h}$, e.g., if $u\geq e$. In the simulations we consider $u\ge e$, which is usually the case in practice.

	For the optimization of the coding schemes, we fix the  generator of the cyclic permutation group to $\pi = (1\;e\;e-1\;\cdots\;2)$ for Schemes~1 and 2 whereas for Scheme~3 we vary $n'$ and assign the submatrices $\bm C_\ell$ as described in \cref{sec:SchemeW}. For Scheme~3, we use $\pi_{\mathsf s} = (1\;\max(n,e)\;\max(n,e)-1\;\cdots\;2)$ and if $n' \geq e$, we use $\pi_{\mathsf c} = (1\;n'\;n'-1\;\cdots\;2)$.
	We then optimize the other parameters for a given privacy level $z$. Particularly, 
		we perform an exhaustive search over all feasible parameter values.	For each set of parameters, unless otherwise stated, we generated $10^4$ instances of the random setup times $\{\lambda_j\}$ and simulated the scheme. We then select the parameters that yield the best expected overall latency  over the $10^4$ runs.
	
\begin{figure}
		\centering
	  	\includegraphics[width=1\columnwidth]{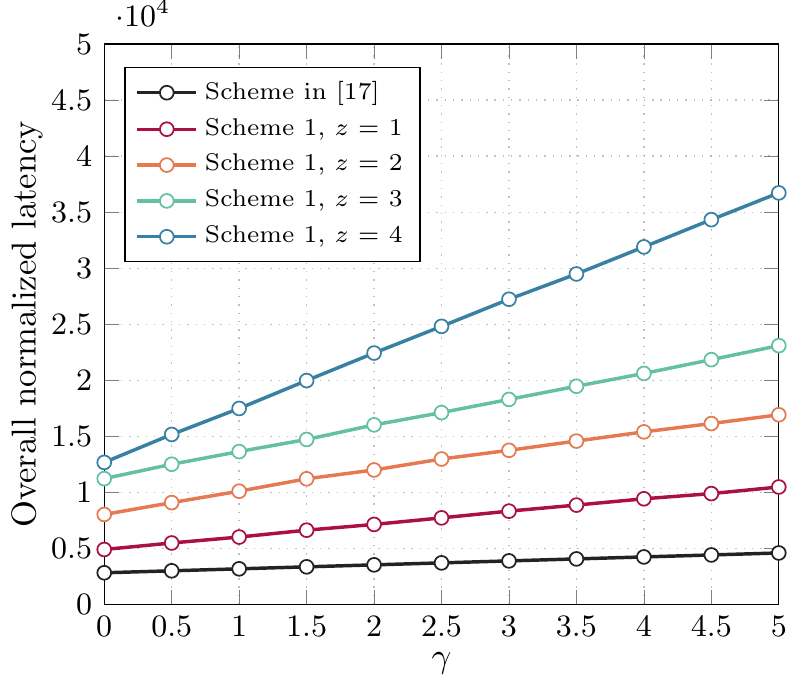}
		\vspace{-2ex}
		\caption{Expected overall normalized latency as a function of $\gamma$ for different privacy levels $z$ of the proposed scheme (Scheme~1) compared to the nonprivate 
	scheme in \cite{Osvaldo}. The parameters are \(\mu = 2/3\), \(\tau = 0.0005\), \(\eta = 0.5\), \(e_{\max}=9\), \(m=600\), \(r=50\), and $\delta = 3$.\label{fig:initial}}
		\end{figure}

	In \cref{fig:initial}, we plot the expected overall latency $\mathbb{E}[\mathsf{L}]$ (given by \eqref{eq:overall_latency}) 	as a function of $\gamma$ for Scheme~1 with different values of $z$ 
	and compare its performance to that of the nonprivate scheme in \cite{Osvaldo}. We remark that in \cite{Osvaldo} both the upload latency and the decoding latency are neglected, while we consider them here. For the scheme in \cite{Osvaldo}, we assume as in \cite{Osvaldo} that the users can broadcast their local data to all ENs simultaneously. However, in general, broadcasting a message to $e$ receivers is more expensive than transmitting a single unicast message to one receiver. As in \cite{Lee}, we assume that broadcasting to $e$ receivers is a factor $\log(e)$ more expensive in terms of latency than a single unicast. Recall that the normalized latency of unicasting $u$ vectors from $\text{GF}(q)^r$ is $\gamma r$. Hence, for the nonprivate scheme in \cite{Osvaldo}, the normalized latency of every user broadcasting its local data to all $e$ ENs is $\mathsf{L}^\mathsf{up}_\mathsf{NP} = \gamma \cdot r\cdot \log(e)$.

	To yield privacy, the proposed scheme involves more communication and computation at the ENs than the nonprivate scheme, as there are multiple shares to be transmitted and computed on instead of a single vector $\x_i$ per user.  As a result, the proposed scheme has a higher latency. As expected, the expected overall latency increases with the privacy level $z$. For $\gamma = 0$, the latency of the private scheme increases by a factor $1.7$, $2.8$, $4.0$, and $4.5$ for $z=1$, $2$,  $3$, and   $4$, respectively, compared to the nonprivate scheme, whereas for $\gamma = 5$ the factors are $2.3$, $3.7$, $5.0$, and $8.0$, respectively. The relative increase in latency increases with $\gamma$ (i.e., increases with the relative communication costs) due to the aforementioned higher communication load of the proposed scheme. We also notice that the proposed scheme does not always utilize all available ENs. For example, for $z=1$ and $\gamma=2.5$, Scheme~1 has the lowest expected overall latency when contacting only $e=6$ ENs. The parameter $e$ influences not only the upload cost, but also the number of submatrices of $\w$, which in turn influences the number of submatrices stored at each EN, $p$, which effects the multiplicity of IRs. This complex interplay of dependencies on $e$ makes it difficult to predict the optimal value of $e$. For example, for $z=1$, the optimal $e$ increases with $\gamma$ (we have $e=8$ for $\gamma\geq4$) whereas for $z=2$, $e$ decreases with $\gamma$ (from $e=9$ for $\gamma \leq 1.5$ to $e=8$ for $\gamma\geq2$).

\begin{figure}
		\centering
	  	\includegraphics[width=1\columnwidth]{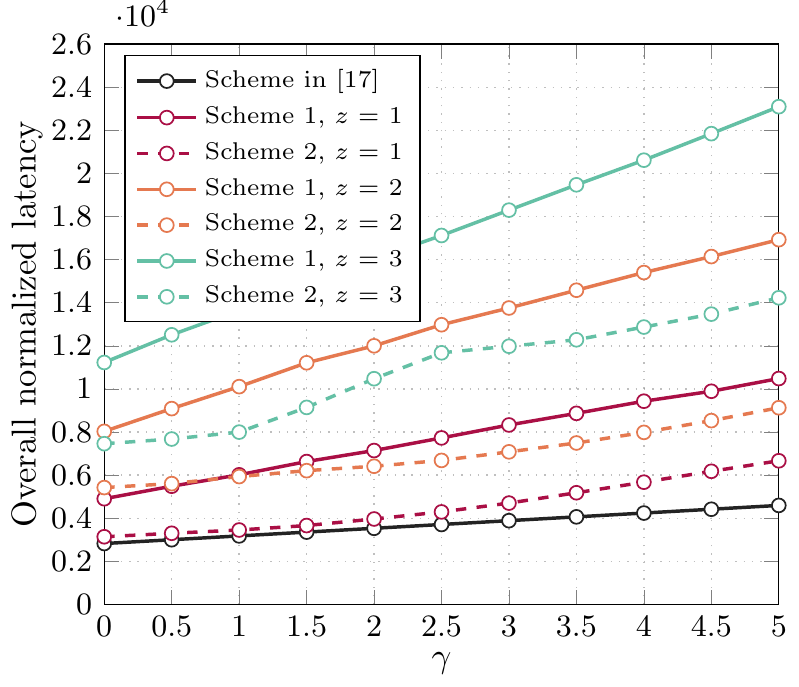}
		\vspace{-2ex}
		\caption{Expected overall normalized latency as a function of $\gamma$ for different privacy levels $z$ of the proposed scheme (Scheme~1) compared to the priority queue variant (Scheme~2) and the nonprivate  scheme in \cite{Osvaldo}. The parameters are \(\mu = 2/3\), \(\tau = 0.0005\), \(\eta = 0.5\), \(e_{\max}=9\), \(m=600\), \(r=50\), and $\delta = 3$.\label{fig:queue}}
		\end{figure}

   In \cref{fig:queue}, we compare the performance of Scheme~1 with that of Scheme~2 and the nonprivate scheme in \cite{Osvaldo}. The use  of a  priority queue reduces the expected overall latency, especially for high values of $\gamma$, i.e., when communication is comparatively expensive. As a result, for $z=1$, Scheme~2 performs similar to the nonprivate scheme, while providing privacy against one honest-but-curious server.

\begin{figure}
		\centering
	  	\includegraphics[width=1\columnwidth]{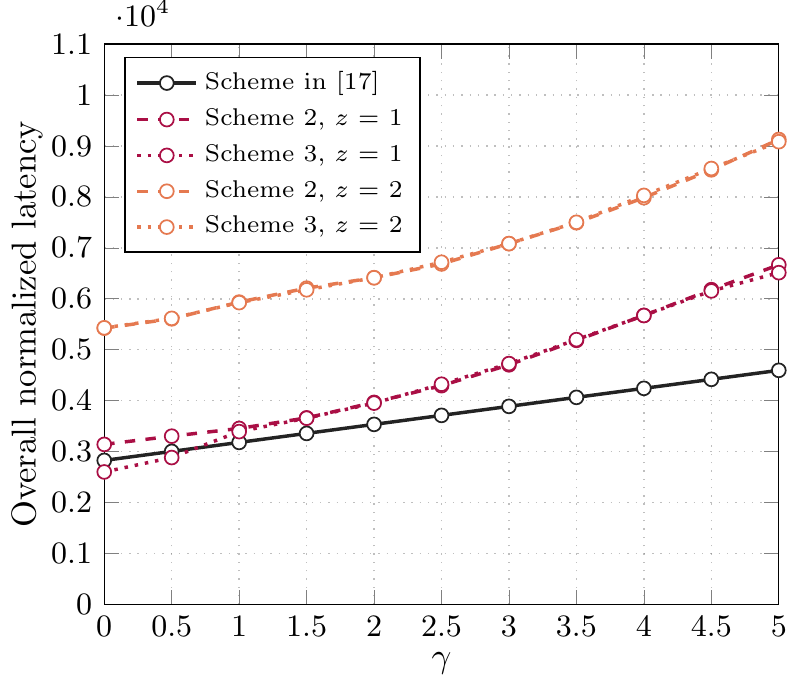}
		\vspace{-2ex}
		\caption{Expected overall normalized latency as a function of $\gamma$ for different privacy levels $z$ of the priority queue variant (Scheme~2), the priority queue with coding on $\w$ variant (Scheme~3), and the nonprivate scheme in \cite{Osvaldo}. The parameters are \(\mu = 2/3\), \(\tau = 0.0005\), \(\eta = 0.5\), \(e_{\max}=9\), \(m=600\), \(r=50\), and $\delta = 3$.\label{fig:queue_codingW}}
		\end{figure}

    In \cref{fig:queue_codingW}, we plot the expected overall latency $\mathbb{E}[\mathsf{L}]$ versus $\gamma$ for Scheme~2, Scheme~3, and the scheme in \cite{Osvaldo}.  The higher flexibility offered by adding redundancy on $\w$ allows to further reduce the expected overall latency with respect to Scheme~2  for low values of $\gamma$, for which the computation times dominate and straggler mitigation is important. Interestingly, this improvement allows the private scheme to outperform the nonprivate scheme for $z=1$. This is explained by the high decoding cost of the scheme in \cite{Osvaldo} compared to the proposed scheme. Indeed, the RS code used in the SSS scheme has very small length and dimension, whereas the MDS code used in \cite{Osvaldo} has much higher length and dimension. For example, for $\gamma=1$ with Scheme~3 and $z=1$ we have $(n',k') = (4,3)$ and for the nonprivate scheme the code length and dimension are in the order of $m$ ($m=600$ in this scenario). Therefore, the nonprivate scheme suffers from higher decoding latency, which significantly penalizes the expected overall latency. For high values of $\gamma$, i.e., when the communication latency becomes more critical, it is beneficial to use as much replication as possible to increase the multiplicities of the IRs to reduce the communication latency in the download. This means that small $n$ and $n'$ are beneficial to reduce the number of distinct IRs. As a consequence, coding on $\w$ brings almost no improvement for high $\gamma$, as we have $n=k$ and $n'=k'$ (i.e., no RS coding) for as low $k$ and $k'$ as possible.

\begin{figure}
		\centering
		\includegraphics[width=1\columnwidth]{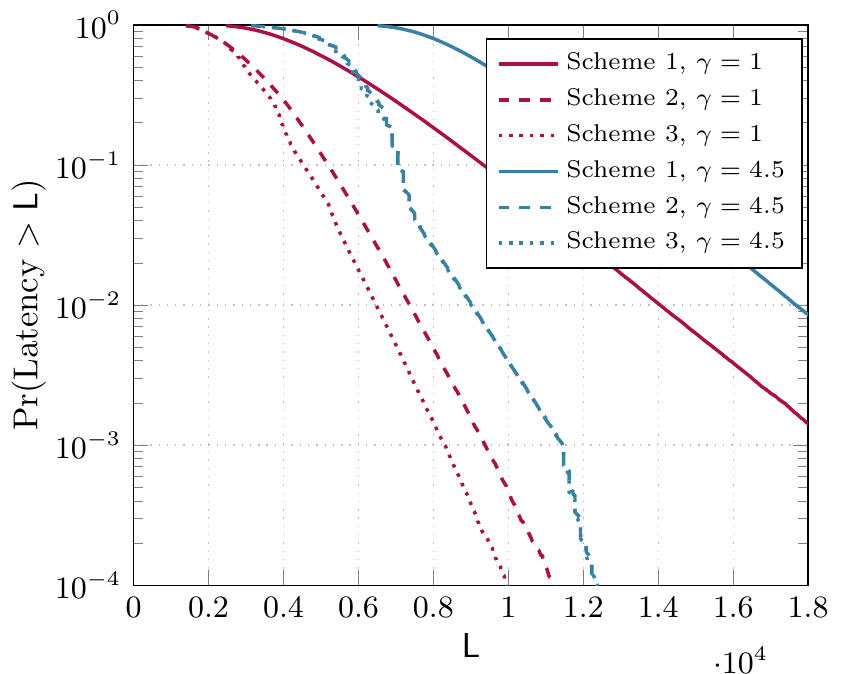}
		\vspace{-2ex}
		\caption{The probability of meeting a given deadline for the private scheme (Scheme~1) and its variants (Schemes~2 and 3) with $z=1$ for different values of $\gamma$.\label{fig:deadline}}
	\end{figure}

	For some applications, the expected  overall latency may not be the most relevant performance metric. %
	In \cref{fig:deadline}, we consider edge computing under a deadline, where we are interested in completing the linear inference within some overall latency. Particularly, 
  we plot the probability that the linear inference is not completed within a deadline $\mathsf{L}$,  for $z=1$ and $\gamma=1$ and $4.5$. To this end, for a given probability, we optimize over (a subset of) the parameters $(\mathscr{C}_{\mathsf{SSS}},\mathscr{C}_{\mathsf{w}},e,\pi,\pi_{\mathsf c},\pi_{\mathsf s},p,t,z)$  for $\pi = (1\;e\;e-1\;\cdots\;2)$, $\pi_{\mathsf c} = (1\;n'\;n'-1\;\cdots\;2)$, $\pi_{\mathsf s} = (1\;\max(n,e)\;\max(n,e)-1\;\cdots\;2)$, and  $z=1$ to minimize $\mathsf{L}$. The number of samples of $\{\lambda_j\}$ is increased to $10^6$ to get reliable results for probabilities down to $10^{-4}$. 
  
  For $\gamma=4.5$ and a deadline $\mathsf{L}=10^4$, the probability of exceeding the deadline is $4.0\cdot10^{-1}$ for Scheme~1, while it decreases to $3.9\cdot 10^{-3}$ for Scheme~2, i.e., two orders of magnitude lower. Introducing coding over $\w$ does not bring further gains. For $\gamma=1$ and a deadline $\mathsf{L}=10^4$, the probability of exceeding the deadline is $7.2\cdot 10^{-2}$ for Scheme~1, while it decreases to $4.5\cdot 10^{-4}$ for Scheme~2. Again, we see an improvement of about two orders of magnitude. Furthermore, for this low value of $\gamma$, introducing coding over $\w$ reduces the probability of not meeting the deadline further to $9.7\cdot10^{-5}$.

%% file: Conclusion.tex
\section{Conclusion}\label{Sec: Conclusion}
	
	We introduced three coded edge computing schemes for linear inference at the network edge that provide privacy against up to $z$ colluding edge servers while minimizing the overall latency encompassing upload, computation, download, and decoding latency. The proposed schemes combine  secret sharing to provide privacy and straggler resiliency, possibly coding over the network model matrix for further straggler mitigation,  and replication of subtasks across edge servers to create cooperation opportunities between edge servers to reduce the download communication latency. 
	Numerical results show that, for a considered  scenario with $9$ edge servers, the proposed scheme yields a $8\%$  latency reduction compared to the nonprivate scheme by  Zhang and Simeone while providing privacy against one honest-but-curious edge server. The privacy level can be enhanced at the expense of a higher latency.

%% file: Appendix.tex
\appendices
\section{Proof of \cref{Th: RecoverSecretComputation}\label{Ap: RecoverSecretComputation}}

		Let $\CSSS$ be the $(n,k)$ RS code used in the SSS scheme. For each \(h\in[n]\), the entries of the rows of \(\bm S^{(h)}\) are code symbols in position \(h\) of codewords from $\CSSS$ pertaining to different users. More precisely, for each user $\mathsf{u}_i$, each row of the matrix \(\bigl(\s^{(1)}_i, \s^{(2)}_i, \ldots,\s^{(n)}_{i}\bigr)\) of all $n$ shares of $\mathsf{u}_i$ is a codeword from $\CSSS$. Since $\CSSS$ is a linear code, each of the $m$ rows of the matrix 
		\begin{align*}
			\w\left(\begin{matrix}
				\s^{(1)}_i, \s^{(2)}_i,\ldots, \s^{(n)}_{i}
			\end{matrix}\right)
		\end{align*}
		is a codeword of $\CSSS$. Furthermore, the messages obtained by decoding these codewords are the rows of
		\begin{align*}
			\left(\w\x_i,\w\bm r_i^{(1)},\ldots,\w\bm r_i^{(k-1)}\right)\, .
		\end{align*}
		Then, decoding the vectors in the set
		\(\{\w\s^{(h)}_i\mid h\in\set{I}\}\)
		gives \(\w\x_i\), and it follows that \(\{\w\bm S^{(h)}\mid h\in\mathcal I\}\) gives  \(\{\w\x_i\}\).

		From the properties of the SSS scheme, it follows  that the mutual information between \(\{\bm S^{(h)}\mid h\in\mathcal J\}\) and $\{\x_i\}$ is zero. Subsequently, from the data processing inequality, it follows  that \(\{\w\bm S^{(h)}\mid h\in\mathcal J\}\) reveals no information about $\{\x_i\}$. %

\section{Proof of \cref{Th: recover}\label{App: recover}}
	
	The proof makes heavy use of combinatorics. For readers unfamiliar with this field, especially the nomenclature of blocks and points, we recommend \cite{Hughes}. We define a map $(x)_e$ that maps an integer $x$ onto the set $[e]$ by successively adding or subtracting $e$ to $x$ until the result lies in $[e]$. For example, for $e=5$,  we have $(3)_5 = 3$, $(-2)_5 = 3$, and $(7)_5 = 2$. In contrast to taking a modulo $e$, we have $(e)_e = e$, whereas $e \mod e = 0$. The rationale for introducing this map instead of the conventional modulo arithmetic is that the indices of matrix rows and columns run from $1$, and not from $0$.
	
	We start by proving the recovery ability. \(\bm I_{\mathsf w}\) is a combinatorial design \(\design{D}\) with \(e\) blocks\textemdash the $e$ sets with entries from the $e$ columns of \(\bm I_{\mathsf w}\)\textemdash  and \(e\) points\textemdash each point is the index $\ell$ pertaining to the submatrix \(\w_\ell\). 
	In particular,  block \(j\) of \(\design{D}\) is \(\set{B}^{(\design{D})}_j=\set{I}^{\mathsf w}_j\).
	Furthermore, each row \(i\) of \(\bm I_{\mathsf s}\) combined with \(\bm I_{\mathsf w}\) represents a combinatorial design \(\design{D}_i\), where its blocks are a permutation \(\pi^{-(i-1)(e-p)}\) of the blocks in \(\design{D}\). 
	More precisely, we have block \(j\) of \(\design{D}_i\) as 
	\begin{align*}
		\set{B}^{(\design{D}_i)}_j=\set{B}^{(\design{D})}_{\pi^{-(i-1)(e-p)}(j)}\, .
	\end{align*}
	Consider \(\bm \Delta^{(\design{D}_i)}\) to be an incidence matrix, of dimensions \(e\times e\), where the incidence relation is between the set of points, \([e]\), and the set of blocks, \(\{\set{B}^{(\design{D}_i)}_j\mid j\in[e]\}\). Then, to prove the recovery ability, we need to show that for
	\begin{align*}
		\bm \Delta=\sum_{i=1}^{\beta+1}\bm \Delta^{(\design{D}_i)}\, , 
	\end{align*}
	we have 
	\begin{align}
		\label{Eq: Th2ProofCompletion}
		\delta_{ij}\geq1,~~\forall i\in[e],~j\in[e]\, ,
	\end{align}
	where \(\delta_{ij}\) is the element in the \(i\)-th row and \(j\)-th column of \(\bm \Delta\).

	We will now show that \cref{Eq: Th2ProofCompletion} holds.
	In the construction of \(\bm I_{\mathsf w}\) and \(\bm I_{\mathsf s}\) in \cref{Eq: Iw,Eq: Is}, respectively, we consider a cyclic permutation group of order \(e\) with elements
	\[\pi^0, \pi, \pi^2,\ldots, \pi^{e-1}\,,\]
	where \(\pi\) is the generator and \(\pi^0\) is the identity element of the group. 
	The set
	\[\{\pi^0(j), \pi(j), \pi^2(j), \ldots, \pi^{e-1}(j)\}=[e]\,,\]
	since \(\pi\) is the generator of the group, and the group is transitive. 
	Let \(\alpha\) be the number of cyclic shifts between two consecutive rows of $\bm I_{\mathsf w}$. 
	Then, 
	\[\pi^i(j)={(j+i(e-\alpha))}_e={(j-i\alpha)}_e\,,\]
	where \(i\in[e]\).
	Note that the blocks of \(\design{D}\) are 
	\begin{align*}
		\mathcal{B}_j^{(\design{D})}=\{\pi^0(j), \pi(j), \ldots,\pi^{p-1}(j)\}\,.
	\end{align*}
	We see that block $j$ consists of \(p\) consecutive permutations of \(j\). 
	Furthermore, for \(d\in[\beta]\),
	\begin{align*}
		\begin{split}
			\pi^{-d(e-p)}(j)&={(j-d(e-\alpha)(e-p))}_e\\
			&={(j-dp\alpha)}_e\,.
		\end{split}
	\end{align*}
	In other words, \(\pi^{-d(e-p)}=\pi^{dp}\). Thus, for some \(j\in[e]\), we have 
	\begin{align*}
		\mathcal{B}_{\pi^{-d(e-p)}(j)}^{(\design{D})}=\{\pi^{dp}(j),\pi^{dp+1}(j),\ldots,\pi^{(d+1)p-1}(j)\}\,,
	\end{align*}
	form which it follows that
	\[\mathcal{B}_{j}^{(\design{D})}\cup\Bigg(\bigcup_{d=1}^{\beta}\mathcal{B}_{\pi^{-d(e-p)}(j)}^{(\design{D})}\Bigg)=[e]\,.\]
	Notice that 
	\(\mathcal{B}_{\pi^{-d(e-p)}(j)}^{(\design{D})}\) is the support of \(\bm \delta^{(\design{D}_{d+1})}_j\), the \(j\)-th column of \(\bm \Delta^{(\design{D}_{d+1})}\). Thus, \cref{Eq: Th2ProofCompletion} holds.
	
	The privacy of the scheme follows straightforwardly. Any $z$ colluding ENs have access to at most $az$ distinct matrices of shares. Since we have $k\geq az+1$, it follows from \cref{Th: RecoverSecretComputation} that the user data privacy is guaranteed.%